\documentclass[letterpaper, twocolumn,superscriptaddress, showpacs,preprintnumbers,amsmath,amssymb]{revtex4}

\usepackage{graphicx}
\usepackage{dcolumn}
\usepackage{bm}

\begin{document}

\title{The many-body localization phase transition} % in one-dimensional interacting hard-core bosons in a random potential}

\author{Arijeet Pal}
\affiliation{Physics Department, Princeton University, Princeton, NJ 08544, USA}

\author{David A. Huse}
\affiliation{Physics Department, Princeton University, Princeton, NJ 08544, USA}

\begin{abstract}
We use exact diagonalization to explore the many-body localization transition in a random-field spin-1/2 chain.
%The model can also be viewed as the spin-1/2 Heisenberg chain with a static random field along the z-direction.
We examine the correlations within each many-body eigenstate,
looking at all high-energy states and thus effectively working at infinite temperature.  For weak random field the
%correlations are consistent with the
eigenstates are thermal, as expected in this nonlocalized, ``ergodic'' phase.  For strong random field the eigenstates are localized,
with only short-range entanglement.  We roughly locate the localization transition and examine
some of its finite-size scaling, %the probability distributions of the correlations,
%asking the question of whether
finding that this quantum phase transition at nonzero temperature might be %more consistent with conventional or with
showing infinite-randomness scaling with a dynamic critical exponent $z\rightarrow\infty$.
\end{abstract}


\maketitle

%\section{Introduction}

\section{Introduction}

As originally proposed in Anderson's seminal paper \cite{pwa58},
an isolated quantum system of many interacting degrees of freedom with quenched disorder may be localized,
and thus generically fail to approach local thermal equilibrium,
even in the limits of long time and large systems, and for energy densities well above the system's ground state.
In the same paper, Anderson also treated the localization of a single particle-like quantum degree of freedom,
and it is this single-particle localization, without interactions,
that has received most of the attention in the half-century since then.
Much more recently, Basko, {\it et al.} \cite{baa} have presented a very thorough study
of many-body localization with interactions at nonzero temperature,
%at low nonzero temperature for interacting fermions,
and the topic is now receiving more attention; see e.g. \cite{gea,OHED,zpp,oph,mm,im,aas,mg,br,fim,gog}.
%The many-body localization

Many-body localization at nonzero temperature is a quantum phase transition that is of very
fundamental interest to both many-body quantum physics and
statistical mechanics:
it is a quantum ``glass transition'' where equilibrium quantum statistical mechanics breaks down.
In the localized phase the system fails to thermally equilibrate.
These fundamental questions about the dynamics of isolated quantum many-body systems are now relevant
to experiments, since such systems can be produced and studied with strongly-interacting ultracold atoms \cite{bdz}.
And they may become relevant for certain systems designed for quantum information processing \cite{sap}.
Also, many-body localization may be underlying some highly nonlinear low-temperature current-voltage
characteristics measured in certain thin films \cite{akla}.

\section{The model}

Many-body localization appears to occur for a wide variety of particle, spin or q-bit models.
Anderson's original proposal was for a spin system \cite{pwa58};
the specific simple model we study here is also a spin model,
namely the Heisenberg spin-1/2 chain with random fields along the $z$-direction \cite{zpp}:
\begin{equation}
H = \sum_{i=1}^L [h_i \hat{S}_i^{z} + J\hat{\vec{S_i}} \cdot \hat{\vec{S}}_{i+1}]~,
\end{equation}
where the static random fields $h_i$ are independent random variables at each site $i$, %with
each with a probability distribution that is uniform in $[-h, h]$.
Except when stated otherwise, we take $J=1$.  The chains are of length $L$ with periodic boundary conditions.
This is one of the simpler models that shows a many-body localization transition.  Since we will be studying the system's behavior by exact diagonalization,
working with this one-dimensional model that has only two states per site %and only nearest-neighbor interactions
allows us to probe longer length scales than would be possible for models on higher-dimensional lattices or with more states per site.
We present evidence that at infinite temperature, $\beta=1/T=0$, and in the thermodynamic limit, $L\rightarrow\infty$,
the many-body localization transition at $h=h_c\cong 3.5 \pm 1.0$ does occur in this model.
The usual arguments that forbid phase transitions at nonzero temperature in one dimension do not apply here,
since they rely on equilibrium statistical mechanics, which is exactly what is failing at the localization transition.
We also present indications that this phase transition might be in an infinite-randomness universality class
with an infinite dynamical critical exponent $z\rightarrow\infty$.

Our model has two global conservation laws: total energy, which is conserved for any isolated quantum system with a time-independent Hamiltonian;
and total $\hat{S^z}$. The latter conservation law is not essential for localization,
and its presence may affect the universality class of the phase transition.
For convenience, we restrict our attention to states with zero total $\hat{S^z}$.
%This restriction does enter in determining finite-size effects, but we see no reason for it to affect the nature of the phase transition.

For simplicity, we consider infinite temperature, where all states are equally probable
(and where the sign of the interaction $J$ does not matter).
The many-body localization transition also occurs at finite temperature;
by working at infinite temperature we remove one parameter from the problem, and use all the eigenstates from
the exact diagonalization (within the zero total $\hat{S^z}$ sector) of each realization of our Hamiltonian.
We see no reason to expect that the nature of the localization transition differs between infinite and finite nonzero temperature,
although it is certainly different at strictly zero temperature \cite{gs}. % is certainly different.
%It is very important to emphasize that what we are studying
Note that this is a quantum phase transition that occurs at nonzero (even infinite) temperature.
Like the more familiar ground-state quantum phase transitions,
this transition is a sharp change in the properties of the many-body eigenstates of the Hamiltonian,
as we discuss below.  But unlike ground-state phase transitions,
the many-body localization transition at nonzero temperature appears to be only a dynamical phase transition
that is invisible in the equilibrium thermodynamics \cite{OHED}.

There are many distinctions between the localized phase at large random field $h>h_c$
and the delocalized phase at $h<h_c$.
We call the latter the ``ergodic'' phase, although precisely how ergodic it is remains to be fully determined \cite{biroli}.
The distinctions between the two phases all are due to differences in the properties of the many-body eigenstates of the Hamiltonian,
which of course enter in determining the dynamics of the isolated system.

In the ergodic phase ($h<h_c$), the many-body eigenstates are thermal \cite{deutsch,sred,tasaki,rdo},
so the isolated quantum system can relax to thermal equilibrium under the dynamics due to its Hamiltonian.
In the thermodynamic limit ($L\rightarrow\infty$), {\it the system thus successfully serves as its own heat bath} in the ergodic phase.
In a thermal eigenstate, the reduced density operator of a finite subsystem converges to the equilibrium thermal distribution for $L\rightarrow\infty$.
%And thus in a many-body eigenstate in the ergodic phase,
Thus the entanglement entropy between a finite subsystem and the remainder of the system is,
for $L\rightarrow\infty$, the thermal equilibrium entropy of the subsystem.
At nonzero temperature, this entanglement entropy is extensive,
proportional to the number of degrees of freedom in the subsystem.

In the many-body localized phase ($h>h_c$), on the other hand, the many-body eigenstates are not thermal \cite{baa}:
the ``Eigenstate Thermalization Hypothesis'' \cite{deutsch,sred,tasaki,rdo} is false in the localized phase.
Thus in the localized phase, the isolated quantum system {\it does not relax to thermal equilibrium} under the dynamics of its Hamiltonian.
The infinite system fails to be a heat bath that can equilibrate itself.
It is a ``glass'' whose local configurations at all times are set by the initial conditions.
Here the eigenstates do not have extensive entanglement, making them accessible to DMRG-like numerical techniques \cite{zpp}.
A limit of the localized phase that is simple is $J=0$ with $h>0$.  Here the spins do not interact,
all that happens dynamically is local Larmor precession of the spins about their local random fields.  No transport
of energy or spin happens, and the many-body eigenstates are simply product states with each spin either ``up'' or ``down''.

Any initial condition can be written as a density matrix in terms of the many-body
eigenstates of the Hamiltonian as $\rho=\sum_{mn}\rho_{mn}|m\rangle\langle n|$.  The eigenstates have different energies,
so as time progresses the off-diagonal density matrix elements $m\neq n$ dephase from the particular phase
relations of the initial condition, while the diagonal elements $\rho_{nn}$ do not change. % with time.
In the ergodic phase for $L\rightarrow\infty$ all the eigenstates are thermal so this
dephasing brings any finite subsystem to thermal equilibrium.  But in the localized phase the eigenstates are all locally
different and athermal, so local information about the initial condition is also stored in the diagonal density
matrix elements, and it is the permanence of this information that in general prevents the isolated quantum system from relaxing to
thermal equilibrium in the localized phase.

Our goals in this paper are (i) to present results in the ergodic and localized phases
that are consistent with the expectations discussed above, and (ii), more importantly, to examine some of the
properties of the many-body eigenstates of our finite-size systems in the vicinity of the localization transition to %see
try to learn about the nature of this phase transition.  Although the many-body localization transition has been
discussed by a few authors, there does not appear to be any proposals for the nature (the universality class) of this phase
transition or for its finite-size scaling properties, other than some very recent initial ideas in Ref. \cite{mg}.
It is our purpose here to investigate these questions,
extending the previous work of Oganesyan and Huse \cite{OHED}, who looked at the
many-body energy-level statistics of a related one-dimensional model.
Since the many-body eigenstates have extensive entanglement on the ergodic side of the transition, it may be that
exact diagonalization (or methods of similar computational ``cost'' \cite{mg})
is the only numerical method that will be able to access the properties of the eigenstates on both sides
of the transition.

%Interestingly, the question of quantum ergodicity is also not very well understood. Ergodicity in classical systems is explained in terms of chaos, where an isolated system thermalizes (equally explores all of phase space) and long-time averages can be replaced by ensemble averages. In comparison, there is no such analogue for isolated quantum systems. The most prominent conjecture at present which seeks to explain thermalization in interacting quantum systems is the \textit{`Eigenstate Thermalization Hypothesis'} (ETH) \cite{Srednicki, Deutsch}. It states that for a generic non-integrable interacting many-body system, the matrix element of any local operator in a particular energy eigenstate is equal to the microcanonical ensemble average at that energy.

%Defining the equilibriation of an isolated quantum system is also a subtle idea. As any system starting in a pure state $|\psi \rangle = \sum_n a_n(0) | n \rangle$ under unitary time evolution evolves into

%\begin{equation}
%| \psi(t) \rangle = \sum_n a_n(0) e^{-i E_n t} | n \rangle
%\end{equation}

%where $| n \rangle$ is the $n^{th}$ energy eigenstate. Under this time evolution the system never equilibrates as it doesn't lose information about its initial conditions because $|a_n(t)|$ is conserved for each eigenstate. So in what sense does an isolated quantum system equilibrate ?

%Next discuss what we use to test ETH in the ergodic phase, statistics of $\langle\hat S_i^z\rangle$? Show and discuss the results.

%Mention that good evidence for the ergodic phase via GOE level statistics was shown for the model studied by OHED.

\section{Does it thermalize?}

As a first simple measure to probe how \emph{thermal} the many-body eigenstates appear to be,
we have looked at the local expectation value of the $z$ component of the spin
\begin{equation}
m^{(n)}_{i\alpha} = \langle n | \hat{S}_i^z | n \rangle_{\alpha}
\end{equation}
at site $i$ in sample $\alpha$ in eigenstate $n$.  For each site in each sample we compare this for eigenstates that are
adjacent in energy, showing the mean value of the difference:
$[|m^{(n)}_{i\alpha}-m^{(n+1)}_{i\alpha}|]$ for various $L$ and $h$ in Fig. 1,
where the eigenstates are labeled with $n$ in order of their energy.  The square brackets denote an average over states, samples and
sites.  The number of samples used in the data shown in this paper
ranges from $10^4$ for $L=8$, to 50 for $L=16$ and some values of $h$.
In our figures we show one-standard-deviation error bars.
Here and in all the data in this paper we restrict
our attention to the many-body eigenstates that are in the middle one-third of the
energy-ordered list of states for their sample.  Thus we look
only at high energy states and avoid states that represent low temperature.  In this energy range, the difference in energy density
between adjacent states $n$ and $(n+1)$ is of order $\sqrt{L }2^{-L}$ and thus exponentially small in $L$ as $L$ is increased.  If the
eigenstates are thermal then adjacent eigenstates represent temperatures that differ only by this exponentially small amount, so the expectation
value of $\hat{S}_i^z$ should be the same in these two states for $L\rightarrow\infty$.  From Fig. 1, one can see that the differences do indeed
appear to be decreasing exponentially with increasing $L$ in the ergodic phase at small $h$, as expected.
[Here and throughout this paper, when we use logarithms, they are base $e$ (``natural'').]
In the localized phase at large $h$, on the other hand, the differences between adjacent
eigenstates remain large as $L$ is increased, confirming that these many-body eigenstates are not thermal.

\begin{figure}[!hbtp]
\includegraphics[height=2.2in,width=3.0in]{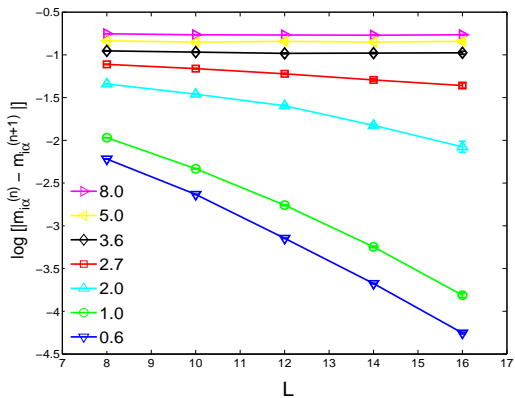}
\caption{(Color online) The logarithm of the mean difference between the local magnetizations in
adjacent eigenstates (see text).  The values of the random field $h$ are indicated in the legend.
In the ergodic phase (small $h$) where the eigenstates are thermal these differences vanish
exponentially in $L$
as $L$ is increased, while they remain large in the localized phase (large $h$).}
\label{Sz_eq}
\end{figure}

Thermalization requires the transport of energy.  In the present model with conserved total $\hat{S^z}$,
it also requires the transport of spin.  To study spin transport on the scale of the sample
size $L$, we consider the relaxation of an initially inhomogeneous spin density:
\begin{equation}
\hat{M}_1 = \sum_j \hat{S}_j^z\exp{(i2\pi j/L)}
\end{equation}
is the longest wavelength Fourier mode of the spin density.  Consider an initial condition
that is at infinite temperature, but with a small
modulation of the spin density in this mode, so the initial density matrix is $\rho_0=(1+\epsilon\hat{M}_1^{\dagger})/Z$,
where $\epsilon$ is infinitesimal, and $Z$ is the partition function.  The initial spin polarization
of this mode is then
\begin{equation}
\langle\hat{M}_1\rangle_0=\sum_n\langle n|\rho_0\hat{M}_1|n\rangle=\frac{\epsilon}{Z}\sum_n\langle n|\hat{M}_1^{\dagger}\hat{M}_1|n\rangle~.
\end{equation}
If we consider a time average over long times, then the long-time averaged density matrix $\rho_{\infty}$ is diagonal
in the basis of the eigenstates of the Hamiltonian, since a generic finite-size system has no degeneracies
and the off-diagonal matrix elements of $\rho$ each time-average to zero.  As a result, the long-time
average of the spin polarization in this mode is
\begin{equation}
\langle\hat{M}_1\rangle_{\infty}=\frac{\epsilon}{Z}\sum_n\langle n|\hat{M}_1^{\dagger}|n\rangle\langle n|\hat{M}_1|n\rangle~.
\end{equation}
Thus for each many-body eigenstate in each sample we can quantify how much it contributes to the initial and to the
long-time averaged polarization.  We then define the fraction of the contribution to the initial
polarization that is dynamic and thus decays away (on average) at long time, as
\begin{equation}
f_{\alpha}^{(n)}=1-\frac{\langle n|\hat{M}_1^{\dagger}|n\rangle\langle n|\hat{M}_1|n\rangle}{\langle n|\hat{M}_1^{\dagger}\hat{M}_1|n\rangle}~.
\end{equation}
In the ergodic phase, the system does thermalize, so the initial polarization does relax
away and $f_{\alpha}^{(n)}\rightarrow 1$ for $L\rightarrow\infty$.  In the localized phase,
on the other hand, there is no long-distance spin transport, so $f_{\alpha}^{(n)}\rightarrow 0$ for $L\rightarrow\infty$.
In Fig. 2 we show the mean values of $f$ for each $L$ vs. $h$.  They show the expected behavior in the
two phases (trending with increasing $L$ towards either 1 or 0), and the phase transition is indicated by the
crossover between large and small $f$ that occurs more and more abruptly as $L$ is increased.

\begin{figure}[!hbtp]
\includegraphics[height=2.2in,width=3.0in]{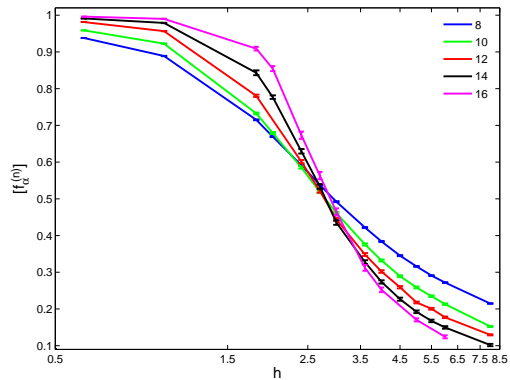}
\caption{(Color online) The fraction of the initial spin polarization that is dynamic (see text).
The sample size $L$ is indicated in the legend. In the ergodic phase (small $h$) the polarization decays substantially under the dynamics,
while in the localized phase (large $h$) the decay is small, and this distinction gets sharper as $L$ increases.}
\label{M1}
\end{figure}

A qualitatively similar finite-size scaling plot also indicating the phase transition is
obtained by examining the many-body eigenenergy spacings as was done in Ref. \cite{OHED}, and is shown as Fig. 3.
We consider the level spacings $\delta^{(n)}_{\alpha}=|E^{(n)}_{\alpha}-E^{(n-1)}_{\alpha}|$, where $E^{(n)}_{\alpha}$ is the
many-body eigenenergy of eigenstate $n$ in sample $\alpha$.  Then we obtain the ratio of adjacent gaps
as $r^{(n)}_{\alpha}=min\{\delta^{(n)}_{\alpha},\delta^{(n+1)}_{\alpha}\}/max\{\delta^{(n)}_{\alpha},\delta^{(n+1)}_{\alpha}\}$, and
average this ratio over states and samples at each $h$ and $L$.  In the ergodic phase, the energy spectrum
has GOE (Gaussian orthogonal ensemble) level statistics and the average value of $r$ converges to
$[r]\cong0.53$ for $L\rightarrow\infty$, while in the localized phase the level statistics are Poisson
and $[r]\rightarrow\cong0.39$. %(fix numbers)  Show plot and discuss.
Note that our model is integrable at $h=0$, so will not show GOE level
statistics in that limit, and this effect is showing up for our smallest $L$ and lowest $h$ in Fig. 3.

\begin{figure}[!hbtp]
\includegraphics[height=2.2in,width=3.0in]{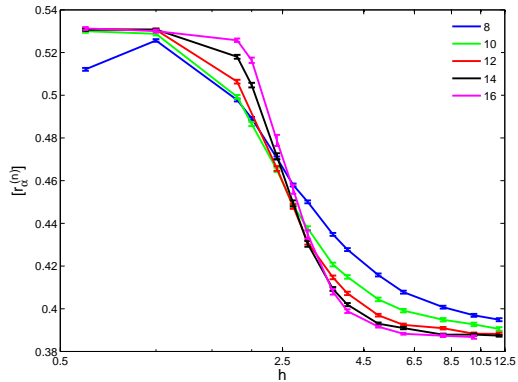}
\caption{(Color online) The ratio of adjacent energy gaps (defined in the text). The sample size $L$ is indicated in the legend. In the ergodic phase, the system has GOE level statistics, while in the localized phase the level statistics are Poisson.}
\label{level spacings}
\end{figure}

The crossings of the curves for different values of $L$ in Figs. 2 and 3 give estimates of
the location $h_c$ of the phase transition.  Both plots show these estimates ``drifting'' towards
larger $h$ as $L$ is increased, with the crossings at the largest $L$ being slightly above $h=3$.
In both cases this ``drifting'' is also towards the localized phase, suggesting the behavior {\it at}
the phase transition is, by these measures, more like the localized phase than it is like the
ergodic phase.

\section{Spatial correlations}

To further explore the finite-size scaling properties of the many-body localization transition in our model, we next look at
spin correlations on length scales of order the length $L$ of our samples.
One of the simplest correlation functions within a many-body eigenstate $|n\rangle$ of the Hamiltonian of sample $\alpha$ is
\begin{equation}
C_{n\alpha}^{zz}(i,j) = \langle n | \hat{S}_i^z \hat{S}_j^z | n \rangle_{\alpha} - \langle n | \hat{S}_i^z | n \rangle_{\alpha} \langle n | \hat{S}_j^z | n \rangle_{\alpha}~.
\end{equation}

\begin{figure}[!hbtp]
\includegraphics[height=2.2in,width=3.0in]{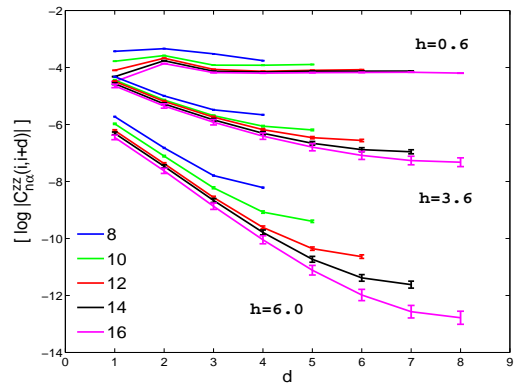}
\caption{(Color online) The spin-spin correlations in the many-body
eigenstates as a function of the distance $d$.  The sample size $L$ is
indicated in the legend.  The correlations decay exponentially with $d$ in
the localized phase ($h=6.0$), while they are independent of $d$ at large $d$
in the ergodic phase ($h=0.6$).  Intermediate behavior at $h=3.6$, which is near the localization
transition, is also shown.}
\label{Czz}
\end{figure}

In Fig. 4 we show the mean value $[\log{|C_{n\alpha}^{zz}(i,i+d)|}]$ as a function of the distance $d$
for representative values of $h$ in the two phases and near
the phase transition.  Data are presented for various $L$.
This correlation function behaves very differently in the two phases:

In the ergodic phase, for large $L$ %and $|i-j|=L/2$
this correlation function
should approach its thermal equilibrium value.  For the states with zero total $\hat S^z$ that we look at, $\langle n | \hat{S}_i^z | n \rangle\cong 0$
in the thermal eigenstates of the ergodic phase.  However, the conservation of total $\hat S^z$ does result in anticorrelations so that
$C_{n\alpha}^{zz}(i,j)\approx -1/(4(L-1))$ for well-separated spins.
These distant spins at sites $i$ and $j$ are entangled and correlated: if spin $i$ is flipped, that quantum
of spin is delocalized and may instead be at any of the other sites, including the most distant one.  These long-range correlations are apparent
in Fig. 4 for $h=0.6$, which is in the ergodic phase.
Note that at large distance the correlations in the ergodic phase become essentially independent of
$d=|i-j|$ at large $L$ and $d$, confirming that the spin flips are indeed delocalized.
Although we only plot the absolute value of the correlations, in fact these correlations
are almost all negative, as expected, in this large $L$ ergodic regime.

In the localized phase, on the other hand, the eigenstates are not thermal and $\langle n | \hat{S}_i^z | n\rangle$ remains nonzero for $L\rightarrow\infty$.
If spin $i$ is flipped, within a single eigenstate that quantum of spin remains localized near site $i$, with its amplitude
for being at site $j$ falling off exponentially with the distance: $C_{n\alpha}^{zz}(i,j)\sim\exp{(-|i-j|/\xi)}$, with $\xi$
the localization length.  In the localized phase the typical correlation and entanglement between two spins $i$ and $j$
thus fall off exponentially with the distance $|i-j|$ (except for $|i-j|$ near $L/2$,
due to the periodic boundary conditions).  This behavior is apparent in Fig. 4
for $h=6.0$, which is in the localized phase and has a localization length
that is less than one lattice spacing.  We note that in the localized phase,
as well as near the phase transition, the long distance
spin correlations $C^{zz}$ are of apparently random sign.

%In Fig. we show the mean value of $\log{|C_{n\alpha}^{zz}(i,j)|}$ ....(?)

The data of Figs. 1-4 show the existence of and some of the differences between the ergodic and localized phases.
We have also looked at %energy level statistics and
entanglement spectra \cite{vadim} of the eigenstates (data not shown), which also support
the robust existence of these two phases. % in this model.
In addition to confirming the existence of these two distinct phases,
we would like to locate and characterize the many-body localization phase transition between them.
However, in the absence of a theory of this transition, the nature of the finite-size scaling is uncertain,
which makes it difficult to draw any strong conclusions from these data with their modest range of $L$.
In studies of ground-state quantum critical points with quenched randomness, very broadly speaking, one first step is to
classify the transitions by whether they are governed (in a renormalization group treatment) by fixed points with
finite or infinite randomness \cite{dsf,ssb}.

\begin{figure}[!hbtp]
\includegraphics[height=2.2in,width=3.0in]{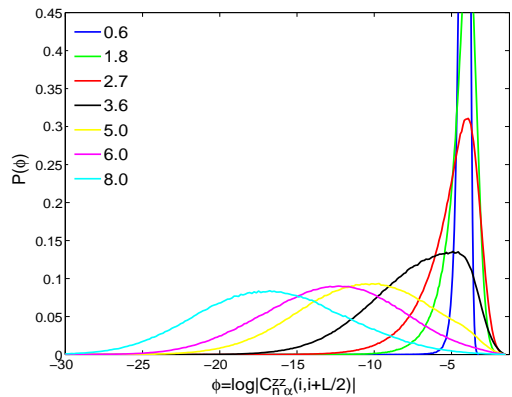}
\caption{(Color online) The probability distributions of the logarithm of the long distance
spin-spin correlation in the many-body
eigenstates for sample size $L=16$ and the values of the random field $h$ is
indicated in the legend.}
\label{Czz_distribution}
\end{figure}

To explore this question for our system, we next look at the probability distributions of the
long distance spin correlations.  For quantum-critical ground states governed by infinite-randomness fixed points, these
probability distributions are found to be very broad \cite{dsf}.  In particular, we look at
\begin{equation}
\phi=\log{|C_{n\alpha}^{zz}(i,i+(L/2))|}~,
\end{equation}
whose probability distributions
for $L=16$ are displayed in Fig. 5 for various values of $h$.  Note the distributions are narrow, as expected, in the
ergodic phase and consistent with log-normal, as expected, in the localized phase.  In between, in the vicinity of the
apparent phase transition, the distributions are quite broad and asymmetric.

\begin{figure}[!hbtp]
\includegraphics[height=2.2in,width=3.0in]{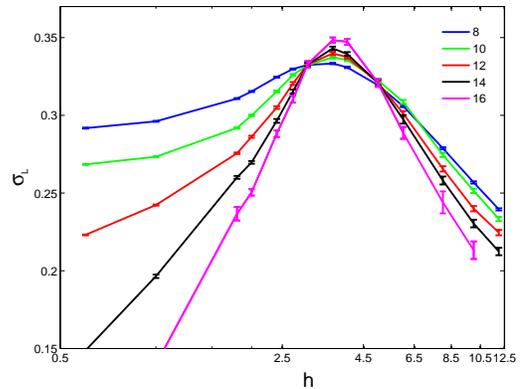}
\caption{(Color online) The scaled width $\sigma$ of the probability distribution of
the logarithm of the long-distance spin correlations (see text).
The legend indicates the sample lengths $L$.  In the ergodic phase
at small $h$ and in the localized phase at large $h$, this width decreases with increasing $L$,
while near the transition it increases.  To produce the one-standard-deviation error bars
shown, we have calculated the $\sigma$ (see text) for each sample by averaging only over sites and
eigenstates within each sample, and then used the sample-to-sample variations of $\sigma$
to estimate the statistical errors.  We have also (data not shown) calculated $\sigma$ by instead averaging
$\phi$ and $\phi^2$ over all samples; this produces scaling behavior for $\sigma$ that is
qualitatively the same as shown here, but with $\sigma$ somewhat larger in the localized
phase and near the phase transition.}
\label{strong_disorder}
\end{figure}

%We also have examined
To construct a dimensionless measure of how these distributions
change shape as $L$ is increased, we divide $\phi$ by its mean, defining $\eta=\phi/[\phi]$.
Then we quantify the width of the probability distribution of $\eta$ by the standard deviation
$\sigma=\sqrt{[\eta^2]-1}$.  % of $\phi$ divided by the mean of $-\phi$, which gives a dimensionless measure of the shape of the probability distribution.
This quantity is shown in Fig. 6 vs. $h$ for the various
values of $L$.  By this measure, in both the ergodic and localized phases the distributions
become narrower as $L$ is increased, as can be seen in Fig. 6.
This happens in the localized phase because although the mean of $-\phi$ grows linearly in $L$, the standard deviation is expected to
grow only $\sim\sqrt{L}$.  Over the small range of $L$ that we can explore, $\sigma$ is found to decrease more
slowly than the expected $L^{-1/2}$ in the localized phase, but it does indeed decrease.

This scaled width $\sigma_L(h)$ of the probability distribution of $\phi$ as a function of the
random field $h$ for each sample size $L$ shows a maximum between the ergodic and localized
phases.  %We take the location of this maximum as a finite-size estimate of the localization
%transition point $h_c$, since this peak in $\sigma$ vs. $h$
%gets higher and sharper as $L$ is increased, suggesting that it is indeed
%associated with the phase transition.  However, one should be cautious, given
%that we only vary $L$ by a factor of two.
%But from Fig. 4, assuming that the maxima do estimate $h_c$, we suggest that $h_c$ is roughly 3 or 4.
In the vicinity of the phase transition, $\sigma$ actually increases as $L$ is increased,
suggesting that its critical value is nonzero, like for quantum-critical ground states that are governed
by an infinite randomness fixed point.  This suggests the possibility that this one-dimensional many-body localization
transition might also be in an infinite-randomness universality class.
%We also note that the crossings of the $\sigma_L(h)$ curves for different $L$'s on the
%higher $h$ side of the peak ``drift'' towards lower $h$ as $L$ is increased.  This
%suggests that these crossings might be used as upper bounds on $h_c$ that converge to the correct
%$h_c$ from above as $L$ is increased, suggesting that $h_c < 4.5$.  This is in contrast to the other such crossings
%shown in other finite-size scaling plots in this paper that appear to converge to $h_c$ from below.
The peak in this plot is close to $h=4$, and is thus suggesting a slightly higher estimate
of $h_c$ than the crossings in Figs. 2 and 3.

\section{Dynamics}

In the study of the spectral and localization properties of noninteracting particles in
finite samples (such as quantum dots), there are two very important energy scales: the
level spacing $\delta$ and the Thouless energy $E_T$.  The Thouless energy is $\hbar$ times the rate of
diffusive relaxation on the scale of the sample.  The diffusive (nonlocalized or ergodic) phase is
where $E_T$ is larger than $\delta$, and for $d$-dimensional samples with $d\geq 3$, the localization transition
occurs when these two energy scales are comparable.  Since the single-particle level spacing in a $d$-dimensional system
of linear size $L$ behaves as $\delta\sim L^{-d}$ and this sets the relaxation time at the localization
transition, the dynamic critical exponent for the single-particle localization transition is $z=d$.

A possibility that we will now investigate is that the many-body localization transition also
occurs when the Thouless energy is of order the many-body level spacing.  Since the many-body level
spacing behaves as $\log{\delta}\sim -L^d$, this corresponds to an infinite dynamic critical
exponent $z\rightarrow\infty$.  Note also that even for our model with $d=1$ this is a stronger divergence of the critical
time scales than occurs at the known infinite-randomness ground-state quantum critical points, where
$\log{\delta}\sim -L^{\psi}$ with $\psi\leq 1/2$.

\begin{figure}[!hbtp]
\includegraphics[height=2.2in,width=3.0in]{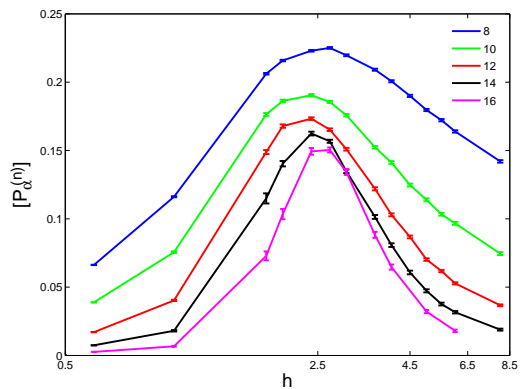}
\caption{(Color online) Contribution to the dynamic part of $\langle \hat{M}_1 \rangle$ from matrix elements between adjacent energy states (see text). In the ergodic and localized phase the contribution is decreasing to zero with increasing sample size. The sample size $L$ is indicated in the legend. The maximum contribution from adjacent states is close to the critical point.}
\label{Thouless_scaling}
\end{figure}

It is important to note that the model (1) we study has two globally conserved quantities; total energy and total $\hat{S}^z$.  Their respective transport times (and hence their corresponding Thouless energy) in the ergodic phase may have different scaling properties close to the critical point. By studying the relaxation of the spin modulation, $\hat M_1$, we are specifically probing the spin transport time which may diverge differently from the energy transport time close to the critical point. Such a possibility has been discussed in the context of zero-temperature metal-insulator transitions \cite{belitz} and may play a role in deciding the universality class of the many-body localization transition.

Naively, the Thouless energy is set by the relaxation rate of the longest-wavelength
spin density modulation, $\hat M_1$.  If the scaling at the many-body localization transition
is such that the Thouless energy is of order the many-body level spacing, then
at the transition a nonzero fraction of the dynamic
part of $\langle\hat M_1\rangle$ should be from its matrix elements between
adjacent energy levels, and this fraction should remain large as $L$ is
increased.  In each sample $\alpha$, the contribution of a given eigenstate $|n\rangle$
to the dynamic part of $\langle\hat M_1\rangle$ is given by
\begin{equation}
(\Delta M_1)_{\alpha}^{(n)}=\langle n|\hat{M}_1^{\dagger}\hat{M}_1|n\rangle - |\langle n|\hat{M}_1|n\rangle|^2~.
\end{equation}
In the ergodic phase, $(\Delta M_1)_{\alpha}^{(n)}$ has significant contributions from matrix elements
with many other eigenstates, and the Thouless energy is a measure of the energy range over which these
contributions occur.  %In the ergodic phase, we expect the Thouless energy to be large compared to the level spacing.
%Hence, the contributions to the dynamic part are from an exponentially large number of neighboring states.
%While at the many-body localization transition if $E_T$ is of order $\delta$ then the number of states contributing to the dynamic part will be of $O(1)$.
To quantify this, we define $Q_{i\alpha}^{(n)}$ as the contribution to the dynamic part of $\langle\hat M_1\rangle$
from the matrix elements between state $n$ and states $n\pm i$: %by the states at distance $i$ from the $n^{th}$ eigenstate in the energy spectrum.
\begin{equation}
Q_{i\alpha}^{(n)} = |\langle n-i|\hat{M}_1|n\rangle|^2 + |\langle n|\hat{M}_1|n+i\rangle|^2
\end{equation}
in sample $\alpha$.  Note that
\begin{equation}
\Sigma_{i\neq0} Q_{i\alpha}^{(n)} = (\Delta M_1)_{\alpha}^{(n)}~.
\end{equation}
We define $P_{\alpha}^{(n)}=Q_{1\alpha}^{(n)}/(\Delta M_1)_{\alpha}^{(n)}$ as the fraction of the longest-wavelength ``diffusive''
 dynamics that is due to interference between adjacent ($i=1$) many-body energy levels.
 Fig. 7 shows this quantity averaged over disorder realizations and states. %This measures the fraction (define more carefully?) of the contribution to the dynamic part of $\langle\hat M_1\rangle$ by the adjacent states.

If at the localization transition the Thouless energy $E_T$ is proportional to the many-body level
spacing $\delta$, then $[P]$ should remain nonzero in the limit $L\rightarrow\infty$.
We do indeed find a strong peak in this fraction
near the many-body localization transition, and that the fraction is large and not
decreasing much as $L$ is increased.  Note that the level spacing decreases by almost
a factor of 4 for every increase of $L$ by two additional spins, so
near the transition the Thouless energy is apparently decreasing by almost the same factor as $L$ is increased.
This seems at least consistent with $E_T\sim\delta$ scaling, and thus dynamic exponent $z\rightarrow\infty$.
In the localized phase, the dynamics is due to spin-moves that are short-range in real space
(probably of order the localization length).  These spin-hops involve pairs of many-body eigenstates
that become far apart (large $i$) for large $L$; this is why $[P]$ drops with increasing $L$
in the localized phase.  Note that the peak in $[P]$ occurs a little below $h=3$.  If one ignores
$L=8$, the location of this peak is apparently drifting to larger $h$ with increasing $L$, consistent
with our other rough estimates of $h_c$.

\begin{figure}[!hbtp]
\includegraphics[height=2.2in,width=3.0in]{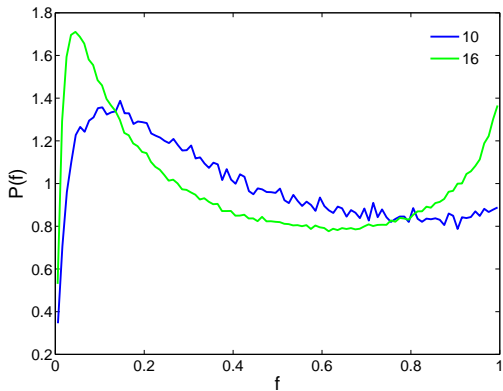}
\caption{(Color online) Probability distribution of the dynamic fraction of $\langle \hat{M}_1 \rangle$ for $L=10$ and 16. Close to the transition for $h=3.0$, the distribution becomes broader and more bimodal with increasing $L$.}
\label{P of f}
\end{figure}

The dynamic fraction $[f_{\alpha}^{ (n)}]$ (Fig. 2) tends to $1$ in the ergodic phase and decreases to $0$ in the localized phase. 
The probability distribution of $f_{\alpha}^{ (n)}$ ($P(f)$) is strongly peaked around $1$ and $0$ in these respective phases.  
At the phase transition, this distribution could either be peaked at the critical $f_c$, broadly distributed, or even bimodal with peaks near both zero and one.  
In Fig. 8, we show $P(f)$ for a disorder strength $h=3.0$ close to the estimated transition, for system sizes $10$ and $16$.  
This distribution $P(f)$ becomes broader and more bimodal with increasing $L$.  
This feature of the distribution is consistent with the indication from Fig. 6 that the critical point may be governed by a strong disorder fixed point. %, based on the distribution of correlations between widely separated spins.

\section{Summary}

This study of the exact many-body eigenstates of our model (1) has demonstrated some of the properties of the ergodic and localized phases. We also find a rough estimate of the localization transition using various different diagnostics. Based on earlier work by one of the authors \cite{OHED}, the many-body energies go from having GOE to Poisson level statistics with increasing disorder. The scaling of the probability distributions of the long-distance spin correlations suggests that the transition might be governed by an infinite-randomness fixed point with dynamic critical exponent $z\rightarrow\infty$. We also study the relaxation of spin modulation under the dynamics of the Hamiltonian. In this case our results are consistent with $E_T \sim\delta$ scaling at criticality, in apparent agreement with our earlier conclusion of $z\rightarrow\infty$ at the transition. These results suggest that efforts to develop a theory of this interesting phase transition should consider the
possibility of a strong disorder renormalization group approach.
Of course, the model we have studied is only one-dimensional, and the behavior of this transition in
higher dimensions might be different in important ways.

\section{Acknowledgement}

We thank Vadim Oganesyan for previous collaborations and for many useful discussions related to this work.
This work was supported by ARO Award W911NF-07-1-0464 with funds from the
DARPA OLE Program and by the NSF
through MRSEC grant DMR-0819860.

\end{document}